\documentclass{elsarticle}
\usepackage{graphicx}
\usepackage{geometry}
\usepackage{amsmath}
\usepackage{subfigure}
\usepackage{hyperref}
\usepackage{array}
\usepackage[normalem]{ulem}
\usepackage{subcaption}
\usepackage[english]{babel}

\title{Novel Instabilities in Counter-Streaming Nonabelian Fluids}
\author[a]{Subramanya Bhat K N
\fnref{fn1}\corref{cor1}}
\ead{subramanyabhatkn@gmail.com}
\author[a]{Amita Das\fnref{fn2}\corref{cor1}}
\ead{amita@iitd.ac.in}
\author[a]{V Ravishankar\fnref{fn3}\corref{cor1}}
\ead{vravi@iitd.ac.in}
\author[b]{Bhooshan Paradkar\fnref{fn4}}
\cortext[cor1]{Corresponding author}
\fntext[fn1]{subramanyabhatkn@gmail.com}
\fntext[fn2]{amita@iitd.ac.in}
\fntext[fn3]{vravi@iitd.ac.in}

\affiliation[a]{organization={Indian Institute of Technology Delhi},
addressline={Hauz Khaz},
postcode={110016},
city={New Delhi},
country={India}}
\affiliation[b]{organization={UM-DAE Centre for Excellence in Basic Sciences, University of Mumbai},
city={Mumbai},
country={India}}

\begin{document}
\begin{abstract}
The dynamics of strongly interacting particles are governed by  Yang-Mills (Y-M) theory, which is a natural generalization of Maxwell Electrodynamics (ED). Its quantized version is known as quantum chromodynamics (QCD) \cite{Gross:1973id, Politzer:1973fx, G.Hooft} and has been very well studied. Classical Y-M theory is proving to be equally interesting because of the central role it plays in describing the physics of quark-gluon plasma (QGP) — which was prevalent in the early universe and is also produced in relativistic
heavy ion collision experiments. This calls for a systematic study of classical Y-M theories. A good insight into classical Y-M dynamics would be best obtained by comparing and contrasting the Y-M results with their ED  counterparts. In this article, a beginning has been made by considering streaming instabilities in Y-M fluids.  We find that in addition to analogues of ED instabilities,  novel nonabelian modes arise, reflecting the inherent nonabelian nature of the interaction. The new modes exhibit propagation/ growth,  with growth rates that can be larger than what we find in ED. Interestingly, we also find a mode that propagates without getting affected by the medium.
\end{abstract}

\maketitle

\section{Introduction}
It is well known that four fundamental interactions essentially govern the dynamics of the universe. Among them, electrodynamics is uniquely placed: the theories of the remaining three interactions are modelled by generalising its central property, viz, gauge invariance. This generalization is quite nontrivial and leads to many novel features such as self-interaction and inherent nonlinearity. Given this,  it is natural to enquire if phases analogous to that of ED plasma exist for these interactions. 
They are ruled out for gravity because it is purely attractive, while it is well nigh impossible to explore them in weak interactions\footnote{ as indicated by the very name,  they are orders of magnitude weaker than electrodynamics (which is itself a hundred times weaker than the strong force) and are also even more short ranged than strong interactions.}. We are thus left with only one candidate, \textit{viz.}, strong interaction for our study. It turns out that it does furnish the analogue of ED plasma.

Quark Gluon Plasma (QGP) \cite{ranjanRaviQGPintroduction} is the exact counterpart of electrodynamic plasma.  Recall that hadrons such as protons, neutrons, and mesons are bound states of quarks or of quark-antiquarks, QGP is their deconfined phase\footnote{There is, however, one fundamental difference. While particles carrying electric charges are experimentally observed, free quarks and gluons that carry the strong charge are not been detected directly in experiments. Thus, deconfinement is similar to metal-insulator transition where the electron states get delocalised within the material, but are still bound to the lattice as a whole.}. 
The early universe, just before hadronization, was in this phase.  More pertinently for observations, the QGP is also produced in relativistic heavy-ion collision experiments \cite{QGP-discovery, ADAMS2005102, RAMELLO200659} in accelerators, and has been a subject of much study.  It is of importance to us that there is good evidence that many of the QGP properties, especially those involving flow,  can be described in the classical language \cite{CristinaHDynamic}. In this context, an \textit{ab initio} study of streaming instabilities could be a useful starting point for studying Y-M dynamics.

There is a large body of earlier works where classical Y-M theory has been studied in the context of QGP \cite{filamentation-pokrovsky, Jithesh-oscillation, Kajantie-plasmons-classical-GT, Kaw-QGP-oscillations,SENGUPTA1999104}. Some examples are thermalization of QGP, hydrodynamic evolution and Debye screening \cite{Bhatt1994-screening, Chandra2009, equilibration-ravi, Debye} in an expanding plasma by employing a Vlasov description, Weibel-like instabilities in nucleon-nucleon collisions and so forth \cite{Chromodynamic-weibel-NN-collision, Hardloop-dynamics-instabilities-prl, Strickand-Thermaization-plasmaInstability, Strickland-Cinstabilities-in-qgp, Strickland2015-pramana}. Almost all of them are tailor-made to address specific issues pertaining to heavy ion collisions, and they borrow heavily from QCD  (or QCD-inspired theories)  and phenomenology.  They are also largely restricted to studying direct ED analogues such as screening lengths and modes which are insensitive to the intrinsic nonabelian nature of the interaction.  In contrast, we propose to initiate an {\it ab initio} study of streaming instability in Y-M fluids. As already mentioned, we expect them to be of great relevance to heavy ion collisions. 

Yet another motivation, though rather academic is that ED streaming instabilities \cite{Pegoraro-weibel, Pegoraro-weibel-space-evolution, Bret, Atul-Amita-EnergyPrinciple, Shukla-Amita-WeibeleBeam} have received a new fillip in laser interaction with overdense plasmas \cite{Amita-PRR, Das2020-RMPP}. High-intensity lasers would introduce non-linearity while Y-M theories are intrinsically nonlinear. A comparison between the two nonlinearities should also be of interest.

The theory governing quark-gluon interactions is what is known as an $SU(3)$ gauge theory. Its classical version would be a theory described in a phase space which is extended to include an additional eight-dimensional internal space. The charges -- called colour, are vectors in the internal space. To avoid unnecessary group-theoretic complications, we study its simpler version -- an $SU(2)$ gauge theory. Here the internal space is only three-dimensional because of which it is intuitively simpler to visualise the nonabelian features of the theory. We perform the analysis entirely in the linear response regime. Yet, we see the emergence of new modes which have no ED analogues, and whose dispersion relations are qualitatively different.

The manuscript has been organized into eight sections. We have deliberately chosen to provide a detailed description of the Y-M system and have provided comparisons with the ED plasma wherever it is pertinent to cater to both the plasma and high-energy physics community. For this purpose in 
 section \ref{section:2-YM}, we introduce the governing equations of the classical Yang-Mills fields. The dynamical equations are adapted to a fluid depiction for the evolution of momenta and colour charge. The equations are, of course, dimensionally consistent using which, in section \ref{section:Dimensional considerations}, we identify normalizations (convenient units) and relevant scales.
 Armed with this we introduce, in section \ref{section:Description of the system}, two counter-streaming YM fluids in equilibrium: the total colour charge densities and momentum are zero. We then perform a linear perturbation analysis. As mentioned, we find that in addition to the conventional ED-like plasma modes, a novel set of growing modes arise, entirely owing to the nonabelian nature of the dynamics. There is yet another propagating mode which is completely blind to the plasma. In sections \ref{section:Abelian-Results},\ref{section:Nonabelian-Results} and \ref{section:compare}, we analyse the results in detail and identify regimes in which the ED-like and the novel modes dominate. These regimes would help us in experimentally verifying Y-M predictions.

\section{Classical Yang-Mills Theory}\label{section:2-YM}
 
Yang-Mills theory \cite{YangMills-original, WongClassicalYM-Isospin, WuYang} generalizes  Maxwell's theory in two interrelated ways. First, the charge (called colour)  is a vector in an internal space and acts as a dynamical variable. Similarly, the colour electric and magnetic fields are also vectors in the internal space. Secondly, gauge transformations, apart from acting on the gauge potentials, also act on the vectors in internal space. In our case, they are just rotations. We say that they transform covariantly under gauge transformations. Since three-dimensional rotations {\it do not} commute, Y-M theory is called nonabelian. As a consequence of the nonabelian dynamics, the fields themselves carry charge and become self-interacting which, in turn,  makes the theory nonlinear. Thus charge conservation should take the flow of charge between the field and matter into account.   Finally, unlike in ED,  gauge potentials are a necessity for the very formulation of the theory. To make these notions clearer, and for the sake of pedagogy, we first present a brief review of YM equations in subsection \ref{section:YM}. As mentioned, they are written for the $SU(2)$ gauge theory.

\subsection{YM dynamics: a brief review}\label{section:YM}
Yang-Mills equations for $SU(2)$ gauge theory are given  by, 
\begin{subequations}
    \label{eq:YM}
    \begin{eqnarray}
        \partial_i \vec{E}_i - g \vec{A}_i \times \vec{E}_i &= &  4\pi \vec{\rho}  \label{eq: YM1} \\
        \epsilon_{ijk}\partial_{j} \vec{B}_k - g \epsilon_{ijk}\vec{A} _{j} \times \vec{B}_{k} -  \frac{1}{c} \partial_0 \vec{E }_{i} -  g  \vec{\phi}  \times \vec{E}_{i}   & = &  \frac{4\pi}{c}\vec{J}_i \label{eq: YM2} \\
        \partial_i \vec{B}_i -  g\vec{A}_i \times \vec{B}_i& = &  0 \label{eq: YM3} \\
        \epsilon_{ijk}\partial_{j} \vec{E}_k - g \epsilon_{ijk}\vec{A} _{j} \times \vec{E}_{k} +\frac{1}{c} \partial_0 \vec{B}_{i} + g  \vec{\phi}  \times \vec{B}_{i}    & = & 0   \label{eq: YM4} 
    \end{eqnarray}
\end{subequations}
Eqs (\ref{eq: YM1})-(\ref{eq: YM4}) merit some explanation. Note the occurrence of both the vector sign and Latin indices (as subscripts). The former indicates that the quantities have three components in the internal colour space. The Latin indices merely represent spatial components. Also note that the gauge potentials $\vec{\phi},~ \vec{A}_i$ (which generalise the usual scalar and vector potentials) appear explicitly in the equations. This is completely unlike what we find in Maxwell's equations. As usual, derivatives with respect to space and time variables are denoted by $\partial_i = \partial/\partial x_i$ and $\partial_0 = \partial/\partial t$ respectively. A la ED, the colour charge and current densities are denoted by $\vec{\rho}$ and $\vec{J}_i$. Similarly, $\vec{E}_i$ and $\vec{B}_i$ represent the colour electric and magnetic fields. Finally, note the appearance of the coupling constant $g$ which is the measure of YM dynamics. If we were to set $g=0$, the equations would collapse to a triplet of uncoupled Maxwell fields.

Equations (\ref{eq: YM1})-(\ref{eq: YM4}) get completed by specifying how the gauge potentials generate the fields. They are given by, 
\begin{subequations}
    \label{eq:EBinPot}
    \begin{eqnarray}
        \vec{E}_i & = & -\partial_i \vec{\phi} - \frac{1}{c}\partial_0 \vec{A}_i + g \vec{A}_i \times \vec{\phi} \label{eq:EinPot}  \\
        \vec{B}_i  & = & \epsilon_{ijk} \partial_j \vec{A}_k  - g \epsilon_{ijk}\vec{A}_j \times \vec{A}_k \label{eq:BinPot}
    \end{eqnarray}
\end{subequations}
Once we employ these definitions, Eqs (\ref{eq: YM3})-(\ref{eq: YM4}) become mathematical identities.

Y-M equations inform us on how the sources generate the fields. To know how the colour charges respond to the fields,
equations of motion need to be added. As mentioned, colour is a dynamical variable. Hence, in addition to the counterpart of the Lorentz force, we have an equation that governs the evolution of the charge vector, and is known as the Wong equation. Explicitly, the equations of motion are
\begin{equation} 
    \label{eq:emom}
    \frac{dP_i}{dt}   =   g \vec{Q}\cdot\left ( \vec{E}_i + \epsilon_{ijk}\frac{V_j}{c}\vec{B}_k\right) 
\end{equation}
\begin{equation}
    \label{eq:ech}
    \frac{d\vec{Q}}{dt} =  gc \vec{Q} \times \left( \vec{\phi} - {\vec{A}_i}\frac{V_i}{c}\right)
\end{equation}
An analogy is helpful here. The coupling constant $g$ is analogous to the gravitational constant $G$, and the charge $\vec{Q}$ is the counterpart of mass $m$. Thus, the charge $\vec{Q}$ couples to the Y-M fields through $g$. In the Maxwellian limit, $g$ cannot be determined independently of $\vec{Q}$. The electrodynamic limit is, therefore, consistently obtained by taking the limits $g \rightarrow 0$, $Q \rightarrow \infty$, keeping $gQ$ fixed. 
Finally,  if the charge and current densities are continuous functions, in the classical regime that we are, it also follows that the number density should also be continuous and must satisfy the continuity equation ($V_i(x_i, t)$ is the velocity field)
 \begin{equation}
    \partial_0 N + \partial_i (N V_i) = 0
    \label{eq:continuity}
\end{equation}
whose inclusion naturally induces a fluid picture of the system.

\subsection{Gauge transformations and gauge choice}
We now describe the gauge transformations that leave the set of Y-M equations and the equations of motion for the particle covariant. The transformations are local, by which we mean that the axis $\hat{n}$ and angle of rotation $\theta$ in the colour space vary continuously with position and time coordinates\footnote{The special case where there is no variation is called a global transformation.}. Let  $R(\hat{n}(\vec{r}, t), \theta (\vec{r}, t))$ be the matrix that denotes the transformation. If we define a vector $\vec{\chi} = \hat{n}\theta$, then,the gauge fields transform as
\begin{subequations}
    \begin{eqnarray}
        \phi^a \rightarrow \phi^{\prime a} &= & R\left(\hat{n}(\vec{r}, t), \theta (\vec{r}, t)\right)^{ab} \phi^b - \frac{1}{gQc} \frac{\partial}{\partial t} \chi^{a}(x_i, t)   \label{eq:GT1} \\
        {A_i}^{a} \rightarrow {A_{i}}^{ \prime a}  & = & R\left(\hat{n}(\vec{r}, t), \theta (\vec{r}, t)\right)^{ab} {A_i}^{b} + \frac{1}{gQ} \frac{\partial}{\partial x_i} \chi^{a}(x_i, t) \label{eq:GT2}
    \end{eqnarray}
\end{subequations}
It is clear from Eqs (\ref{eq:GT1}-\ref{eq:GT2}) that the gauge potentials are {\it not} vectors in the colour space. For, even if they were vanishing in one coordinate system, they pick up nonzero values once a gauge transformation is performed. The only exception occurs when the transformation is local. Be that as it may, the colour charge, electric field and the colour magnetic field have a covariant character. Thus,  under a gauge transformation,  we have 
\begin{subequations}
    \begin{eqnarray}
        Q^a \rightarrow Q^{\prime a} &= & R\left(\hat{n}(\vec{r}, t), \theta (\vec{r}, t)\right)^{ab} Q^b \\
        E_i^a \rightarrow E_i^{\prime a} &= & R\left(\hat{n}(\vec{r}, t), \theta (\vec{r}, t)\right)^{ab} E_i^b \\
        B_i^a \rightarrow B_i^{\prime a} &= & R\left(\hat{n}(\vec{r}, t), \theta (\vec{r}, t)\right)^{ab} B_i^b
    \end{eqnarray}
\end{subequations}
The covariance property of the charges and the fields allows us to choose a convenient gauge to work with. As in ED, one may employ a variety of gauges such as Coulomb and Lorentz gauges\footnote{The complete formalism presented in the section may be derived from a Lagrangian formalism.}. Yet another very useful gauge, which we shall employ in this paper is the temporal gauge where we set $\vec{\phi} =0$.


\section{Dimensional considerations}\label{section:Dimensional considerations}
The quantum version of Y-M theory employs only one coupling constant $g$,  and does not introduce the charge $Q$. That is made possible because the quantum theory has an additional natural constant $\hbar$. The classical theory does not have this luxury and we are perforce required to introduce $Q$. For this reason, it is instructive to perform a simple dimensional analysis so that we may avoid erroneous comparisons with ED quantities. 
The dimensions of most quantities are quite similar to the {\it{Gaussian}} units employed in ED. It follows from Eqs (\ref{eq:EBinPot}) and (\ref{eq:emom}) that 
 
\begin{eqnarray}
    [A, \phi ]   & = &  M^{1/2}L^{1/2}T^{-1} \nonumber \\ 
    \left[E, B \right]  & =  &  M^{1/2}L^{-1/2}T^{-1} \nonumber \\
    \left[g \right] & = & M^{-1/2}L^{-3/2}T \nonumber \\
    \left[Q \right] & = &  ML^{3}T^{-2} \nonumber
\end{eqnarray}
As mentioned, $Q$ does not have the dimension of electric charge $e$, but it is the product $gQ$ that has the right dimensions. In fact, $[Q] = [e^2]$ and $[g] = [e^{-1}]$
  
\subsection{ Scales and normalizations}\label{sec:normalization}
We first identify the fundamental scales that are characteristic of Y-M dynamics. As observed, even in the absence of 
sources $\vec{\rho}$ and  $\vec{J}_i$, the fields are self-interacting, with a strength $g$. Combining it with $c$, we get 
the first scale which has the dimension of angular momentum and is given by
\begin{equation}
    H = \frac{1}{g^2c}
\end{equation}
If we consider purely the matter sector, we get a fundamental length and time scale, that are given by
\begin{equation}
    \ell = \frac{Q}{mc^2};~ \tau = \frac{\ell}{c}
    \label{eq:YMscales}
\end{equation}
Medium-dependent scales emerge when we consider a medium, Y-M plasma in our case. If its equilibrium number density is $n_0$, new time and length scales emerge which are, apart from multiplicative factors, the familiar plasma frequency and the skin depth. They are given by 
\begin{equation}
    \omega_p = \sqrt{\frac{4 \pi n_0g^2Q^2}{m}};~ d = \frac{c}{\omega_p}
    \label{eq:EDscales}
\end{equation} 
Finally, if there is a characteristic speed scale $U$  also associated with the plasma, we get yet another length scale given by
\begin{equation}
    \kappa_{YM} = \left(\frac{8mc \omega_p^2}{QU}\right)^{1/3}
\end{equation}
For instance, $U$ could correspond to the streaming flow velocity of the colour fluid considered in the present study. It is convenient to express all other quantities in units of scale from the above list, chosen depending on the context. In this paper, time scales are expressed in units of $\omega_p$, lengths in units of the skin depth, gauge potentials in units of $a = \frac{gQ}{mc^2}$ and fields in units of $f= \frac{gQ}{mc\omega_p}$. 
In setting up these scales, we have implicitly assumed that we are working in a non-relativistic regime. Thus mass is treated as a constant.

\subsection{Units employed}
In this paper, we shall employ the scales mentioned in the previous section \ref{sec:normalization} to express fields and kinematical variables in appropriate units, and thus deal with dimensionless normalised quantities. The basic units are listed in Table \ref{tab:table 1}.  
\begin{table}[h!]
    \centering
        \begin{tabular}{|m{6cm}|m{4cm}|}
            \multicolumn{2}{c}{\textbf{Table 1}} \\ [1.0ex]
            \hline
            \textbf{Normalised variables} & \textbf{Normalizing factor} \\ [1ex] \hline
            $t_N$ (time)  &  $\omega_p^{-1}$\\ [1ex]\hline 
            $x_N$ (length) & $d $ \\ [1ex]\hline
            $V_N$  (velocity) & $c $ \\ [1ex]\hline
            $N_N$ (number density) & $n_0 $ \\ [1ex]\hline
            $\vec{A}_{i,N},\vec{\phi}_N$ (vector and scalar potential) &  $\frac{mc^2}{gQ}$ \\ [2ex]\hline
            $\vec{E}_{i,N}, \vec{B}_{i,N}$ (electric and magnetic fields)  &  $\frac{mc \omega_p}{gQ} $ \\ [2ex]\hline
        \end{tabular}
        \caption{Units employed in this paper}
        \label{tab:table 1}
\end{table}
In these units, the basic equations obtain an elegant form involving only dimensionless variables. We list them below in order, by suppressing the subscript $N$ (used in Table 1 to indicate normalized variables). 
The equations defining the colour electric and magnetic fields read 
\begin{subequations}
    \label{eq:n-EBinPot}
    \begin{eqnarray}
        \vec{E}_i & =  & -\partial_i \vec{\phi} - \partial_0 \vec{A}_i + \alpha_{YM} \vec{A}_i \times \vec{\phi}
        \label{eq:n-Epot} \\
        \vec{B}_i  & = &  \epsilon_{ijk} \partial_j \vec{A}_k  - \alpha_{YM} \epsilon_{ijk}\vec{A}_j \times \vec{A}_k  
        \label{eq:n-Bpot}
    \end{eqnarray}
\end{subequations}
Next, the YM equations acquire the form
 \begin{subequations}
    \label{eqs:n-YM}
    \begin{eqnarray}
        \frac{\partial \vec{E}_{i}}{\partial x_i} - \alpha_{YM} \vec{A}_{i} \times \vec{E}_{i}  &=& \vec{\rho}  
        \label{eq:n-divE}\\
        \epsilon_{ijk} \frac{\partial \vec{B}_k}{\partial x_j}  - \frac{\partial \vec{E}_i}{\partial t} - \alpha_{YM} \vec{\phi} \times \vec{E}_{i} -\alpha_{YM} \epsilon_{ijk} \left( \vec{A}_j \times \vec{B}_{k}\right) &=& \vec{J}_{i}   
        \label{eq:n-curlB}\\
        \frac{\partial \vec{B}_{i}}{\partial x_i}  -\alpha_{YM} \vec{A}_{i} \times \vec{B}_{i} &=& 0
        \label{eq:n-divB}\\
        \epsilon_{ijk} \frac{\partial \vec{E}_k}{\partial x_j} + \frac{\partial \vec{B}_i}{\partial t} + \alpha_{YM} \vec{\phi} \times \vec{B}_{i} - \alpha_{YM} \epsilon_{ijk} \left( \vec{A}_j \times \vec{E}_{k}\right)  &=& 0
        \label{eq:n-curlE}
    \end{eqnarray}
\end{subequations}
The equations of motion   acquire the form
\begin{subequations}
    \label{eqs:n-eom}
    \begin{eqnarray}
        \frac{dV_{i}}{dt}  &=&  \hat{e}_{Q}\cdot\left ( \vec{E}_i + \epsilon_{ijk}V_{j}\vec{B}_k\right)  
        \label{eq:n-mom}\\
        \frac{d \hat{e}_{Q}}{dt}  &=&  \alpha_{YM}\hat{e}_{Q}\times\left ( \vec{\phi} - V_{i}\vec{A}_i\right) 
        \label{eq:n-wong}
    \end{eqnarray}
\end{subequations}   
The dimensionless constant appearing in equations (\ref{eq:n-EBinPot})-(\ref{eqs:n-eom}) is given by
\begin{equation}
    \alpha_{YM} = mc^3/\omega_{p} Q
    \label{eq:alpha-YM-def}
\end{equation}
It should be noted that $\alpha_{YM}$ essentially corresponds to the ratio of intrinsic length/time scales arising from the nonabelian  Y-M matter sector (Eq.(\ref{eq:YMscales})) with that associated with ED plasma-like scales (Eq.(\ref{eq:EDscales})). As expected, as 
$\alpha_{YM} \rightarrow 0$, the equations reduce to three decoupled abelian ED equations. 
Thus, the value of $\alpha_{YM}$ measures the strength of the nonabelian contribution. 

Finally, the normalized continuity equation for the number density reads
\begin{equation}
    \frac{\partial N}{\partial t}  = - \frac{\partial}{\partial x_i} (N V_{i})  
    \label{eq:n-cont}
\end{equation}

\section{Description of the system}\label{section:Description of the system}
The stage is now set to define the system. At the outset, we set up the notations for the coordinate systems. Rectangular Cartesian coordinate systems will be employed both in the position space and the internal colour space. The orthonormal basis in the position space will be denoted by the set $\{\hat{x}, \hat{y}, \hat{z}\}$. Similarly, we denote the orthonormal basis in the colour space by the set $\{\hat{e}_{1},\hat{e}_{2},\hat{e}_{3}\}$. 

We consider two counter-streaming Yang-Mills fluids in equilibrium. Each stream is made of particles of the same mass $m$, have the same equilibrium number density $n_0$, and move with respective velocities $\pm U\hat{x}$. The corresponding charges of the particles are then given by
$\vec{Q} = \pm Q\hat{e}_1$. Accordingly, the total charge density vanishes everywhere, but there is a finite uniform current $\vec{J}_i = g n_0 Q U\vec{e}_1(2, 0, 0)$ in the system. 

Our object of study is the behaviour of the system under small perturbations, which will be denoted by calligraphic letters. In short,
\begin{eqnarray}
& & \vec{E}_i = \vec{E}_{eq,i} + \mathcal{\vec{E}}_i; \hspace{0.2in}\vec{B}_i = \vec{B}_{eq,i} +\mathcal{\vec{B}}_i; \hspace{0.2in} \vec{A}_i = \vec{A}_{eq,i} + \mathcal{\vec{A}}_i \hspace{0.2in} \vec{Q}_{s} = \vec{Q}_{eq,S} + \vec{q}_{S}  \nonumber \\
& & N_s = N_{eq,S} + n_{S};\hspace{0.1in} V_{i,S} = V_{eq,i,S} + v_{i,S}; \hspace{0.1in}  \vec{\rho} = \vec{\rho}_{eq} + \vec{\varrho}; \hspace{0.1in} \vec{J}_{i} = \vec{J}_{eq,i} +\vec{j}_i
\label{eq:def-lin-pert}
\end{eqnarray}
where $S $ denotes the label for the two-fluid species   $A$ and  $B$  which move with $U \hat{x}$ and $-U \hat{x}$ respectively. As mentioned,  only the colour current $\vec{J}_i$  of the overall system comprising these two fluids is non-vanishing at equilibrium. 
We employ translational invariance along the $y-z$ plane and thus set the equilibrium fields $\vec{E}_{eq,i} = \vec{B}_{eq,i} =0$.
\begin{figure}[h!]
    \centering
    \includegraphics[width=0.70\textwidth]{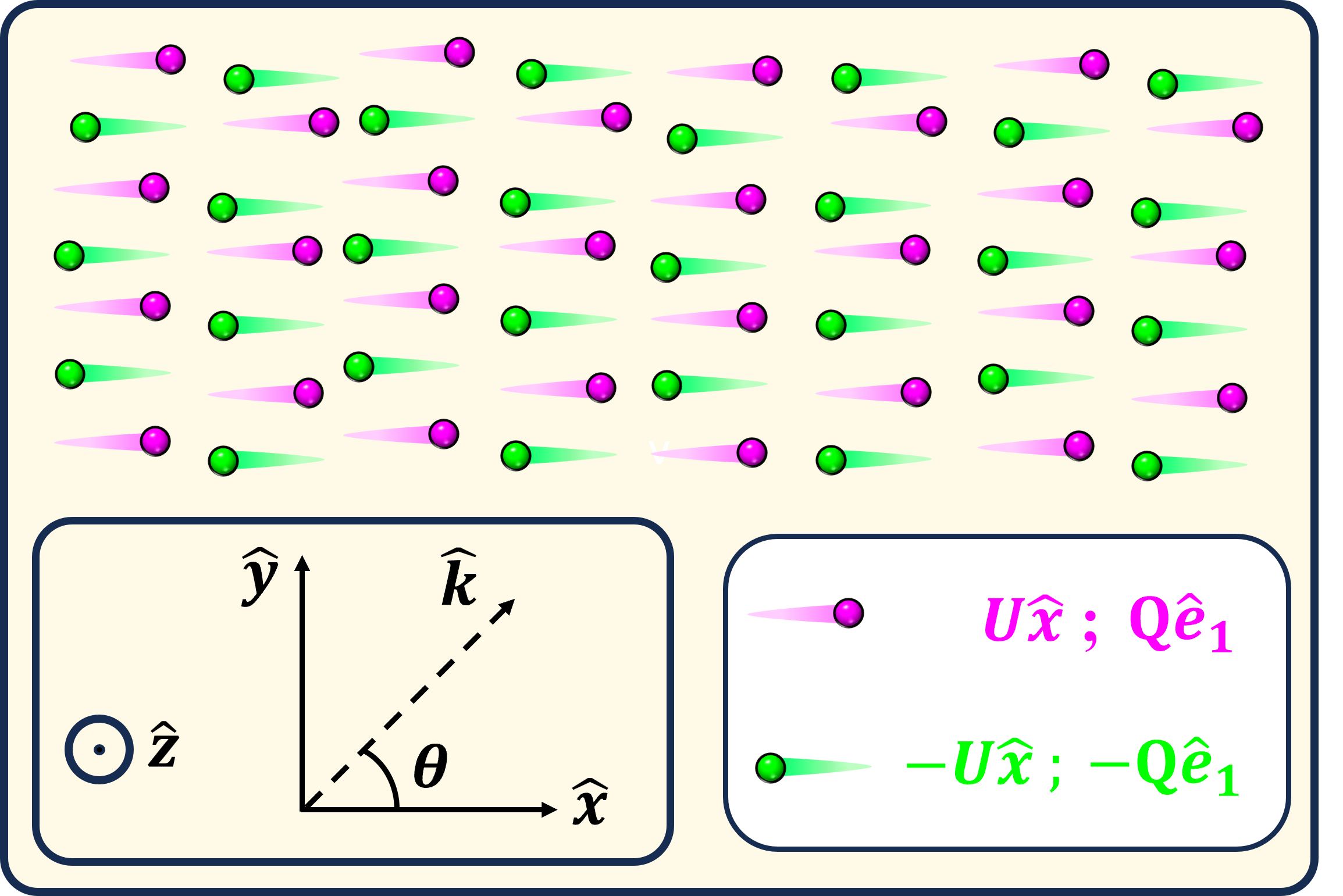}
    \caption{Schematic illustration of the equilibrium configuration of the Y-M fluids.}
    \label{The Geometry}
\end{figure}
It is now of interest to seek the linear response of the system when the system is perturbed from its equilibrium configuration slightly. We systematically linearise the equations of Y-M fields, the momentum equation, Wong, and continuity equations and eliminate all the variables in terms of the colour electric field. We restrict to the 2-D space variation of the perturbed fields in the $x-y$ plane. The linearized set of equations is  Fourier analyzed in space and time. The wave vector $\vec{k}$ thus lies in the 2-D $x-y$ plane.    In Fig.1 the schematic geometry of the configuration has been shown. We now discuss and characterize the modes.

At the outset, we order the set of chromo-electric field components, further organized as  subsets,  as follows:
\begin{equation}
\left\{\{{\cal E}_z^{(1)}, {\cal E}_z^{(2)}, {\cal  E}_z^{(3)}\}, \{{\cal E}_x^{(1)}, {\cal E}_y^{(1)}\}, {\{\cal E}_x^{(2)}, {\cal E}_y^{(2)},
{\cal E}_x^{(3)}, {\cal E}_y^{(3)}\}\right\}
\end{equation}
On performing the self-consistent calculation using Eqs (\ref{eq:n-Epot})-(\ref{eq:n-cont}), we may arrange the response of the fluids, in the basis given above,  in the form of a  $9\times 9$ response matrix. The general form of the matrix is given by,

\begin{equation}
   {\cal R} =  \begin{pmatrix}
        \mathcal{R}_{3 \times 3}& \textbf{0}& \textbf{0}\\
        \textbf{0}&\mathcal{R}_{2 \times 2} & \textbf{0}\\
         \textbf{0}& \textbf{0}&\mathcal{R}_{4 \times 4}
    \end{pmatrix}  
    \label{eq:mat-gen}
\end{equation}
where, as is clear, the first block corresponds to the response to the electric field fluctuations in the $\hat{z}$ direction which, as we may recall, is normal to both the $\vec{k}$ and $  \vec{U}$. The second block gives the electric field fluctuations in the $x-y$ plane and along the internal direction $\hat{e}_1$.  These two blocks give rise to dispersion relations which are only abelian in nature and their form is essentially the same as ED modes. The third block gives rise to novel nonabelian modes and holds a number of interesting features.   They correspond to fluctuations in the $x-y$ spatial plane and $\hat{e}_2-\hat{e}_3$ plane in the internal space. The dispersion relations in each sector are obtained by demanding that the determinant of the corresponding matrix $\mathcal{R}_{i \times i}$ vanishes.
We first discuss the abelian modes in the next section \ref{section:Abelian-Results}. 

\section{The Abelian modes}\label{section:Abelian-Results}
\subsection{Modes with  $\hat{z}$ polarization }
We start with the topmost block matrix $\mathcal{R}_{3 \times 3}$ in (\ref{eq:mat-gen}). It has the following explicit form
\begin{equation}
 \mathcal{R}_{3 \times 3} =    \begin{pmatrix}
        \frac{2}{\omega^2-k^2} -1 & 0&0\\
        0&-1& 0\\
         0& 0&-1
    \end{pmatrix}  
    \label{eq:disp-mat-z}
\end{equation}
This provides a dispersion relation of $\omega^2 = k^2 +2$, for 
a colour wave propagating in the $x-y$ plane, which is linearly polarized along the $z$ direction in space and in the internal space it is along $\hat{e}_1$.   The other two directions in the colour space are insensitive to the plasma. Note that this chromodynamical mode is essentially the same as the one we obtain in the context of electrodynamic equal mass plasma. 

\subsection{Abelian modes with other orientation of Electric field} 
We now consider the sector corresponding to $\mathcal{R}_{2 \times 2}$. For this case, the colour electric field fluctuations lie in the  $x-y$ plane and along $\hat{e}_1$ direction in internal space. This internal space direction corresponds to the equilibrium current flow direction.  The response is abelian in this case as the  form of the   sub-matrix $\mathcal{R}_{2\times 2}$ has no 
dependence on  $\alpha_{YM}$ and is given by
\begin{equation}
\mathcal{R}_{2 \times 2} =  \begin{pmatrix}
        M_1 -1  & M_2 \\
        M_3 & M_4 -1
    \end{pmatrix}
\end{equation}
where the matrix elements $M_i$ have the following  complicated expressions, 
\begin{subequations}
\begin{eqnarray}
    M_1 &=& \frac{2}{\omega^2 - k_y^2}\left(\frac{\omega^4 + \omega^2 U^2k^2 + U^4k_x^2 k_y^2}{(\omega^2 - U^2 k_x^2 )^2} \right)  \\
    M_2 &=& -\beta_y \left(\frac{2 \omega^2 U^2 + 2 U^4 k_x^2}{\left(\omega^2 - U^2 k_x^2 \right)^2} - 1\right)  \\
    M_3 &=& -\beta_x \left(\frac{2 U^2 }{\left(\omega^2 - U^2 k_x^2 \right)} - 1\right) \\
    M_4 &=& \frac{2}{\omega^2 - k_x^2}
\end{eqnarray}
\end{subequations}
where,
\begin{subequations}
    \label{eq:beta}
    \begin{eqnarray}
        \beta_x &=& \frac{-k_x k_y}{\omega^2 - k_x^2}\\
        \beta_y &=&\frac{-k_x k_y}{\omega^2 - k_y^2}
    \end{eqnarray}
\end{subequations}
The dispersion relations are inferred from the algebraic condition 
\begin{equation}
    (M_1-1)(M_4-1) - M_2 M_3 = 0
\end{equation}
\noindent
 The dispersion relation in this case provides the modes that are identical to their ED counterparts which exhibit two-stream instability and Weibel filamentation. We shall briefly illustrate this by considering the two limits,  {\it viz.}, $\theta =0$, and $\theta = \pi/2$, that is when the wave vector is parallel and perpendicular to the direction of the flow.
For the case when $\theta = 0$, we  obtain the following form of the dispersion relation:
\begin{equation}
    \omega^4 - \omega^2 \left( 2k^2 U^2 +2\right) + k^4 U^4 - 2k^2 U^2 = 0
    \label{eq:disp-2stream}
\end{equation}
It is clear that we will obtain a negative value of $\omega^2$ for all values of the wavenumber which satisfies the relationship  $k^2 < 2/U^2 $. Note that this is,  in fact,  the same condition as that of the two-stream instability in the ED plasma. When  $\theta = \pi/2$,  at which $k_x = 0$ and $k_y = k$, 
the dispersion relation is given by 
\begin{equation}
      \omega^4 - \omega^2\left(2 + k^2 \right) - 2k^2 U^2 = 0  
\end{equation}
the solution for which is given by
\begin{equation}
    \omega^2 = \frac{1}{2}\left((2+k^2) \pm \sqrt{(2 +k^2)^2 + 8 k^2 U^2} \right)
    \label{eq:two-stream}
\end{equation}
It is thus clear that since   $k$ and $U$ are both non-vanishing, there is always an unstable  branch with  
$\omega^2 < 0$. This unstable mode is similar to that of the filamentation mode in the context of ED plasma. Thus the modes for which the electric field vector is directed along the $\hat{e}_1$ direction in the colour space (same direction as that of the equilibrium charge colour),  
behave in the same manner as ED plasma.

\section{Novel nonabelian modes}\label{section:Nonabelian-Results}
One of the major endeavours in the study of chromodynamics is to identify phenomena which unmistakably signal the nonabelian features so that the theory may be vindicated. While corrections to the abelian terms are no doubt important, signatures -- whose very existence would be ruled out in a purely abelian theory are more valuable. This is particularly true of QGP which is produced in heavy ion collisions. The system that we are studying does show such signatures.

As made clear in Eq (\ref{eq:mat-gen}),  in this case, the electric field fluctuations lie in the  $ x - y$ plane in real space and in  $\hat{e}_2 - \hat{e}_3$ plane, in the colour space. The submatrix 
$\mathcal{R}_{4\times 4}$, which governs the response  has the form
\begin{equation}
\label{R_4}
  \mathcal{R}_{4 \times 4}=   \begin{pmatrix}
        -1 & \beta_y & -i \Tilde{p} & 0 \\
        \beta_x & -1 & 0 & 0 \\
        i \Tilde{p} & 0 & -1 & \beta_y\\
        0 & 0 & \beta_x & -1\\
    \end{pmatrix}
\end{equation}
where
\begin{equation}
    \Tilde{p} = \frac{2 \alpha_{YM}  U^3 k_x}{\left(\omega^2 - k_y^2\right) \left(\omega^2 - U^2 k_x^2\right)}
    \label{eq:p-tilde}
\end{equation}
 The structure of $\mathcal{R}_{4 \times 4}$ immediately establishes the identities
 \begin{equation}
    \left(\omega^2 - k_x^2\right) \mathcal{\Vec{E}}_y^{(2,3)} = -k_xk_y\mathcal{\Vec{E}}_x^{(2,3)}
    \label{eq:ex-ey-rel}
\end{equation}
because of which it is sufficient to study the dispersion relations and their modes with the simpler equation 
\begin{equation}
  \tilde{\mathcal{R}}_4=   \begin{pmatrix}
        -1 +\beta_x\beta_y & -i \Tilde{p}  \\
        i \Tilde{p}  & -1 +\beta_x \beta_y\\
    \end{pmatrix}\begin{pmatrix} 
    \mathcal{E}_x^{(2)}\\ \mathcal{E}_x^{(3)}\\
    \end{pmatrix} = 0
    \label{eq:R4-tilde}
\end{equation}
The resulting dispersion condition is 
\begin{equation}
    \tilde{p} = \pm\left(1 - \beta_x \beta_y \right)
    \label{eq:disp-gen-cond}
\end{equation}
First, we notice that the eigenmodes of Eq (\ref{eq:R4-tilde}) are universal: they are independent of the dispersion relation and are given by 
\begin{equation}
    \mathcal{E}_{R,L} = \mathcal{E}_x^{(2)} \pm i \mathcal{E}_x^{(3)}
    \label{eq:pols}
\end{equation}
We now look at the detailed structure of Eq (\ref{eq:disp-gen-cond}) which 
leads to the quartic  polynomial equation in $\omega^2$, 
\begin{subequations}
    \label{eq:w-gen}
    \begin{eqnarray}
        \left(\omega^2 - k_y^2 \right)\left\{\omega^6 - \omega^4\left(U^2k_x^2 + k^2 \right) + \omega^2\left(U^2k_x^2[k_x^2+k^2 ] - 2\alpha_{YM}U^3k_x \right) + 2\alpha_{YM}U^3k_x^3 \right \} = 0 \\
        \left(\omega^2 - k_y^2 \right)\left\{\omega^6 - \omega^4\left(U^2k_x^2 + k^2 \right) + \omega^2\left(U^2k_x^2 [k_x^2+k^2] + 2\alpha_{YM}U^3k_x \right)  - 2\alpha_{YM}U^3k_x^3 \right\} = 0
    \end{eqnarray}
\end{subequations}
corresponding to the two respective choices for the sign of $\tilde{p}$ in Eq (\ref{eq:disp-gen-cond}). 
Here the first factor   $(\omega^2 - k_y^2)$ gives a trivial mode. We now analyze the second factor which is within the curly braces and is a third-degree polynomial in $\omega^2$. 

We first consider the two specific cases of $\theta = 0$ and $\theta = \pi/2$. This corresponds to having  $\vec{k}$ parallel and perpendicular to $\vec{U}$ respectively.
\subsection{The specific cases ($\theta = 0$ and $\theta = \pi/2$)}
When $\Vec{k}\parallel\Vec{U}$, Eq (\ref{eq:w-gen}) reduces to
\begin{equation}
    \left(\omega^2 - k_x^2 \right)\left\{\omega^4 - \omega^2U^2k_x^2 \mp 2\alpha_{YM}U^3k_x \right\} = 0
    \label{eq:w-theta0}
\end{equation}
The root  $\left(\omega^2 =k_x^2 \right)$ corresponds  to a trivial solution. The second factor within curly braces leads to 
\begin{equation}
    \omega^2 = \frac{1}{2} \left( k_x^2U^2 \pm \sqrt{k_x^4 U^4 \pm \alpha_{YM} 8 U^3 k_x }\right)
    \label{eq:YM-disp-T0}
\end{equation}
The four solutions for $\omega^2$ merit some discussion.  Let us label the respective solutions as $\omega_{\pm \mp}$ in the same order of signs as they occur in Eq (\ref{eq:YM-disp-T0}). Then the modes belonging to $\omega_{++}$ and $\omega_{-+}$ share the common colour polarization in internal space and are given by 
\begin{equation}
    \mathcal{E}_R = \mathcal{E}_x^{(2)} + i\mathcal{E}_x^{(3)}
\end{equation}
at their respective eigenfrequencies. Similarly, the modes belonging to $\omega_{+-}$ and $\omega_{--}$ have the complementary colour mode
\begin{equation}
    \mathcal{E}_L = \mathcal{E}_x^{(2)} - i\mathcal{E}_x^{(3)}
\end{equation}
at their respective eigenfrequencies.
It should be noted that in the limit  $\alpha_{YM} \rightarrow 0$, both $\omega_{- +}, ~\omega_{--}  \rightarrow 0$. On the other hand,  $\omega_{+ +}, ~\omega_{+-}  \rightarrow \pm Uk$ and correspond to the Doppler-shifted frequencies for the two fluids streaming with $\pm U$.

When $\alpha_{YM}$ is finite, a number of features emerge. They are naturally sensitive to the two relative signs. It is clear that $\omega_{++}$ always gives a propagating mode and $\omega_{-+}$ always gives growing modes which, for large $k$ are proportional to $1/k^{\frac{3}{2}}$. The other two modes, $\omega_{+-} ~\text{and} ~\omega_{--}$ have both propagation and growth, the latter being conditional\footnote{Growth in one mode is naturally accompanied by damping in its partner mode. They are not of interest to us.} on the inequality $ k^3 < \kappa_{YM}^3  \equiv {8\alpha_{YM}}/{U}$, beyond which it vanishes. 

Note that the growths are governed by a new medium-dependent nonabelian scale, \textit{viz}, $\kappa_{YM}$. Their detailed features are presented in Table \ref{tab:Property}. It may be seen in Table \ref{tab:Property} that $\mathcal{E}_R$ can give unconditional growth while $\mathcal{E}_L$ can only give unconditional propagation. This asymmetric situation can be understood by noting that the initial choice for the net colour current is parallel to $\hat{e}_1$ even though the net colour charge density vanishes.  

\begin{table}[h!]
    \centering
    \begin{tabular}{|m{1.8cm}|m{1.5cm}|m{4cm}|m{5.5cm}|}
        \multicolumn{4}{c}{\textbf{Table 2}} \\ [1.0ex]
        \hline
        \textbf{Solution}& \textbf{Mode} & \textbf{Nature of the Mode} & \textbf{Condition for Instability/}\newline \textbf{Propagation} \\ [1.5ex]
        \hline
        $\omega_{++}$ & $\mathcal{E}_R$ & Purely Propagating & None \\ [1.5ex]
        \hline
        $\omega_{-+}$ & $\mathcal{E}_R$ & Purely Growing & None \\ [1.5ex]
        \hline
        $\omega_{+-}$ & $\mathcal{E}_L$ & Conditional Growth \newline+ Propagation & $k < \kappa_{YM}$, Growth + Propagation.\newline $k > \kappa_{YM}$, Propagation \\ [1.8ex]
        \hline
        $\omega_{--}$& $\mathcal{E}_L$  &  Conditional Growth \newline+ Propagation  &$k < \kappa_{YM}$, Growth + Propagation. \newline $k > \kappa_{YM}$, Propagation  \\ [1.8ex]
        \hline
    \end{tabular}
    \caption{Brief summary of the defining features of the modes described by Eq.(\ref{eq:YM-disp-T0})}
    \label{tab:Property}
\end{table}

\begin{figure}[h!]
    \centering
    \includegraphics[width=0.80\textwidth]{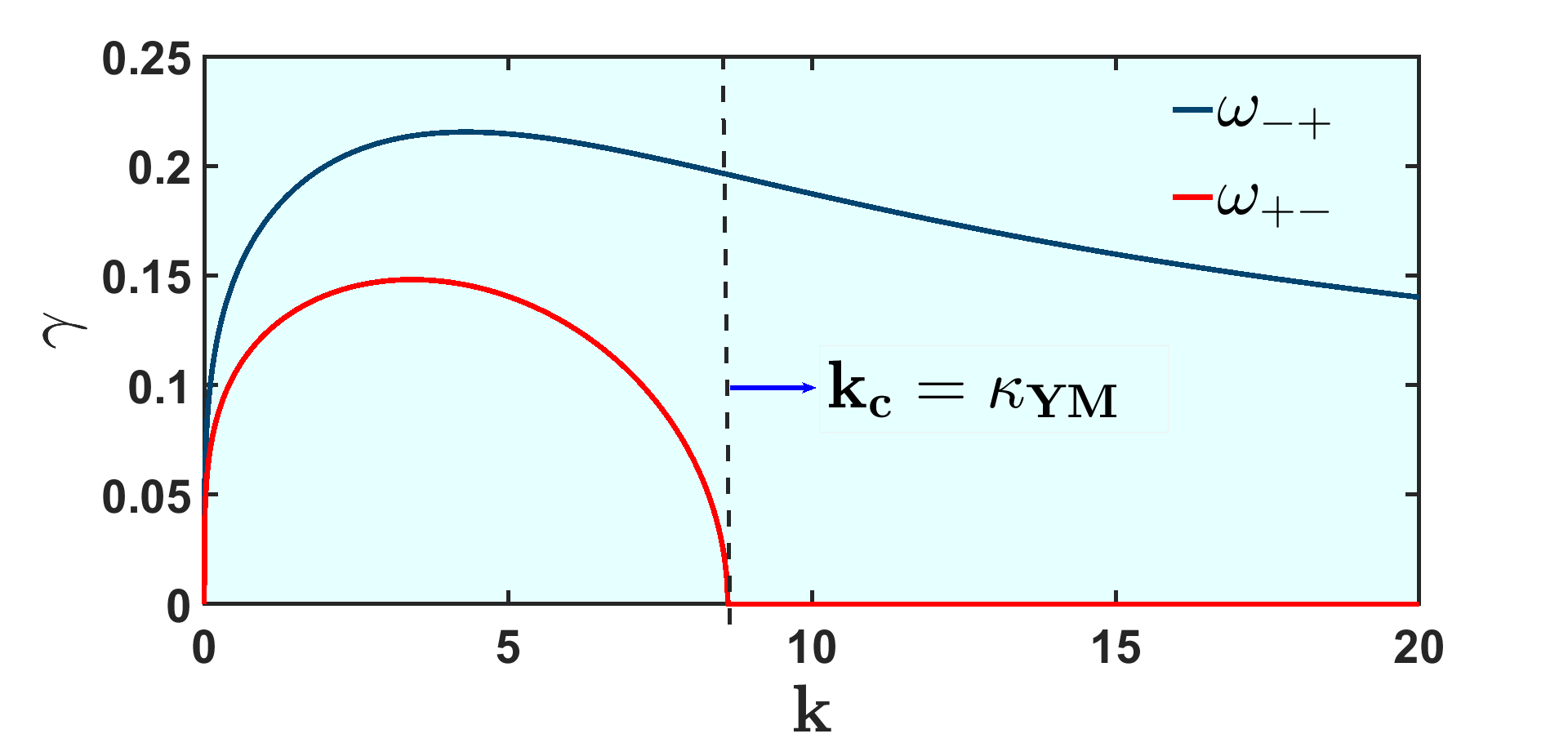}
    \caption{Behaviour of the growth rates for modes $\omega_{-+}$ and $\omega_{+ -}$  at $\theta = 0$ ($U = 0.2, ~and ~\alpha_{YM} = 3.0$). Note that the growth rate for $\omega_{+ -}$ terminates at $k = \kappa_{YM}$ while that of $\omega_{-+}$ persists for all $k$.}
    \label{fig:Growth-YM-theta0}
\end{figure}

For a quantitative illustration, we choose the values $(\alpha_{YM} =3, U = 0.2)$ and plot both the growth rates 
 $\omega_{-+}$ and  $\omega_{+-}$ in Fig \ref{fig:Growth-YM-theta0}. It should be noted that for this particular choice of parameters, the growth rate of $\omega_{-+}$ is higher than that of $\omega_{--} ~\text{or} ~ \omega_{+-}$ for all values of $k$. 
This completes the description of the modes at $\theta = 0$. 
For  $\theta = \pi/2$  (i.e. when $k_x =0$) we do not find any non-trivial nonabelian solutions.  

\subsection{The general case of  $0 < \theta < \pi/2$}
For finite values of  $\theta$  between $0$ and $\pi/2$, 
we have  both  $k_x = k \cos \theta$ and $k_y = k \sin\theta$ as finite. 
The factor in the curly braces of Eqs. (\ref{eq:w-gen}) is then 
solved numerically. It is noted that as one increases the value of  $\theta$ from zero, the growth rate of the modes generally follows the same trend as has been observed at  $\theta =0$. 

\begin{figure}[h!]
    \centering
    \includegraphics[width=0.80\textwidth]{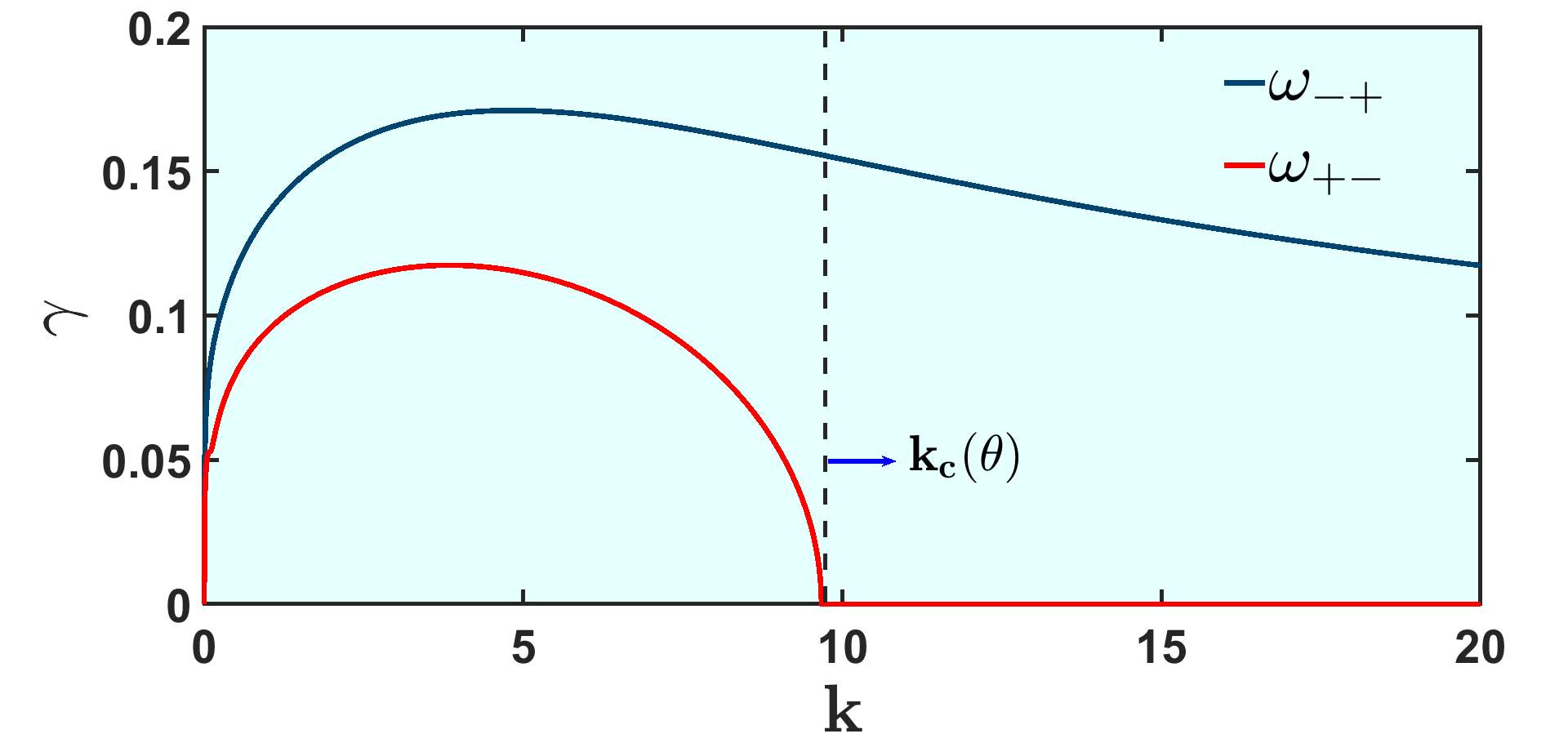}
    \caption {Behaviour of the growth rates of $\omega_{-+}$ and $\omega_{+-}$ as functions of $k$ at $\theta = \pi/4$ ($U = 0.2, \alpha_{YM} = 3.0$).  Note that the growth rate for $\omega_{+ -}$ terminates at $k = k_c$ while that of $\omega_{-+}$ persists for all $k$.}
    \label{fig:Growth-YM-theta45}
\end{figure}

\begin{figure}[h!]
    \centering
    \includegraphics[width=0.80\textwidth]{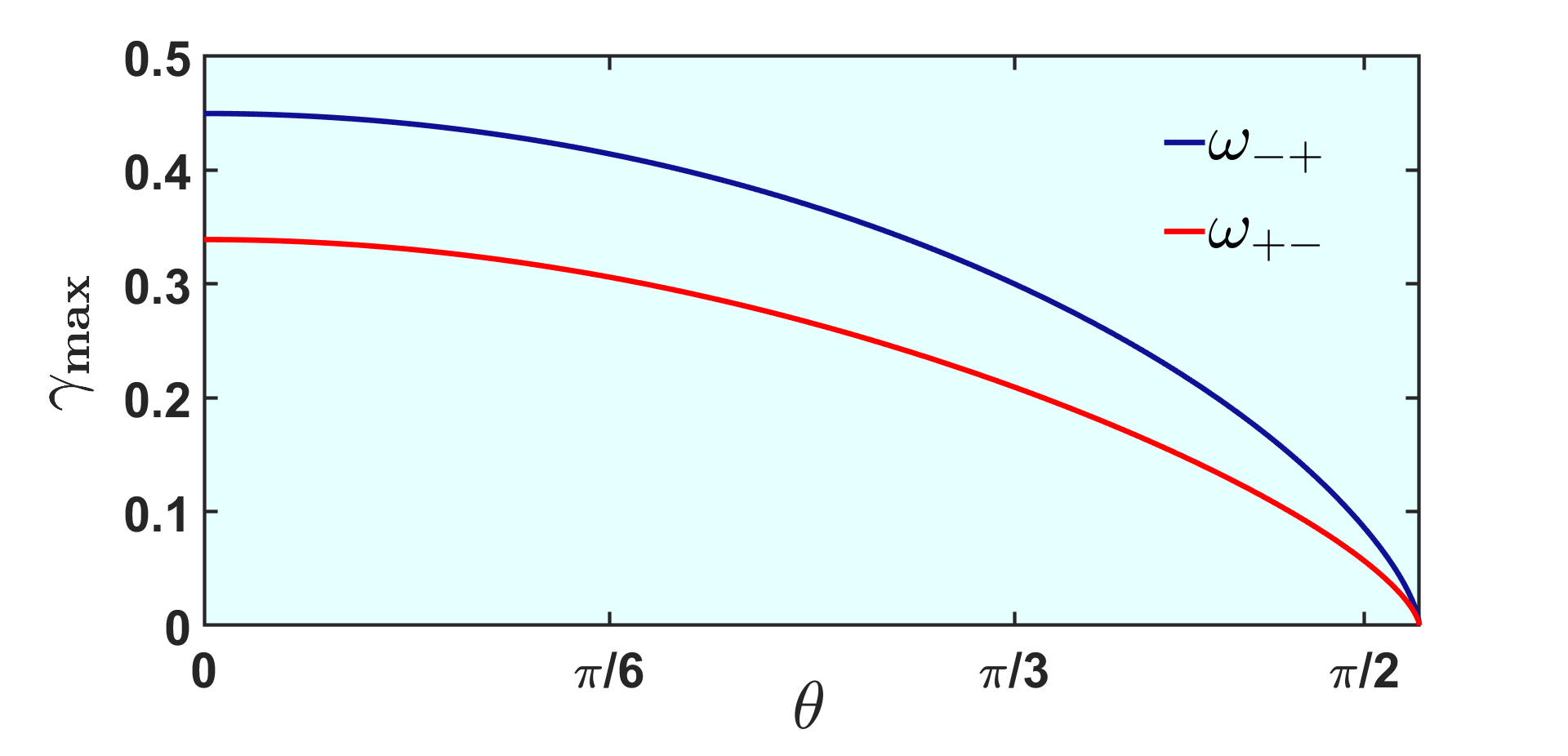}
    \caption{Behaviour of the maximum growth rates $\gamma_{max}$ of the modes $\omega_{-+}$ and $\omega_{+-}$ ($U = 0.2$ and $\alpha_{YM} = 3.0$) as functions of $\theta$. Note that $\gamma_{max} \rightarrow 0$ as angle $\theta \rightarrow \pi/2$.}
    \label{fig:gammaMax-vs-theta}
\end{figure}

However, the maximum growth rates  $(\gamma_{max})$  of both the unstable modes decrease with increasing value of $\theta$. Furthermore, the cut-off wavenumber at which the growth rate of the modes  
$\omega_{+-}$ and $\omega_{--}$ vanish,  increases with $\theta$. These features are illustrated in figures \ref{fig:Growth-YM-theta45}, \ref{fig:gammaMax-vs-theta}. 
The cut-off wavenumber as a function of $\theta$,  ($k_{c} (\theta)$)  has the form of $\kappa_{YM} f(\theta, U)$. Interestingly, $\kappa_{YM}$ is itself composed of three scales which may expressed as
\begin{equation}
    \kappa_{YM}^3  \propto \left(\frac{mc^2}{Q} \right) \left(\frac{\omega_p}{c} \right) \left( \frac{\omega_p}{U} \right)
\end{equation}
which are respectively the fundamental YM scale, the skin depth of the plasma,  and a configuration-dependent scale. Thus, the novel nonabelian modes that we encounter carry the full richness of the nonabelian dynamics which would not be captured in quantities such as the Debye screening length. 


\begin{figure}[h!]
    \centering
    \includegraphics[width=0.80\textwidth]{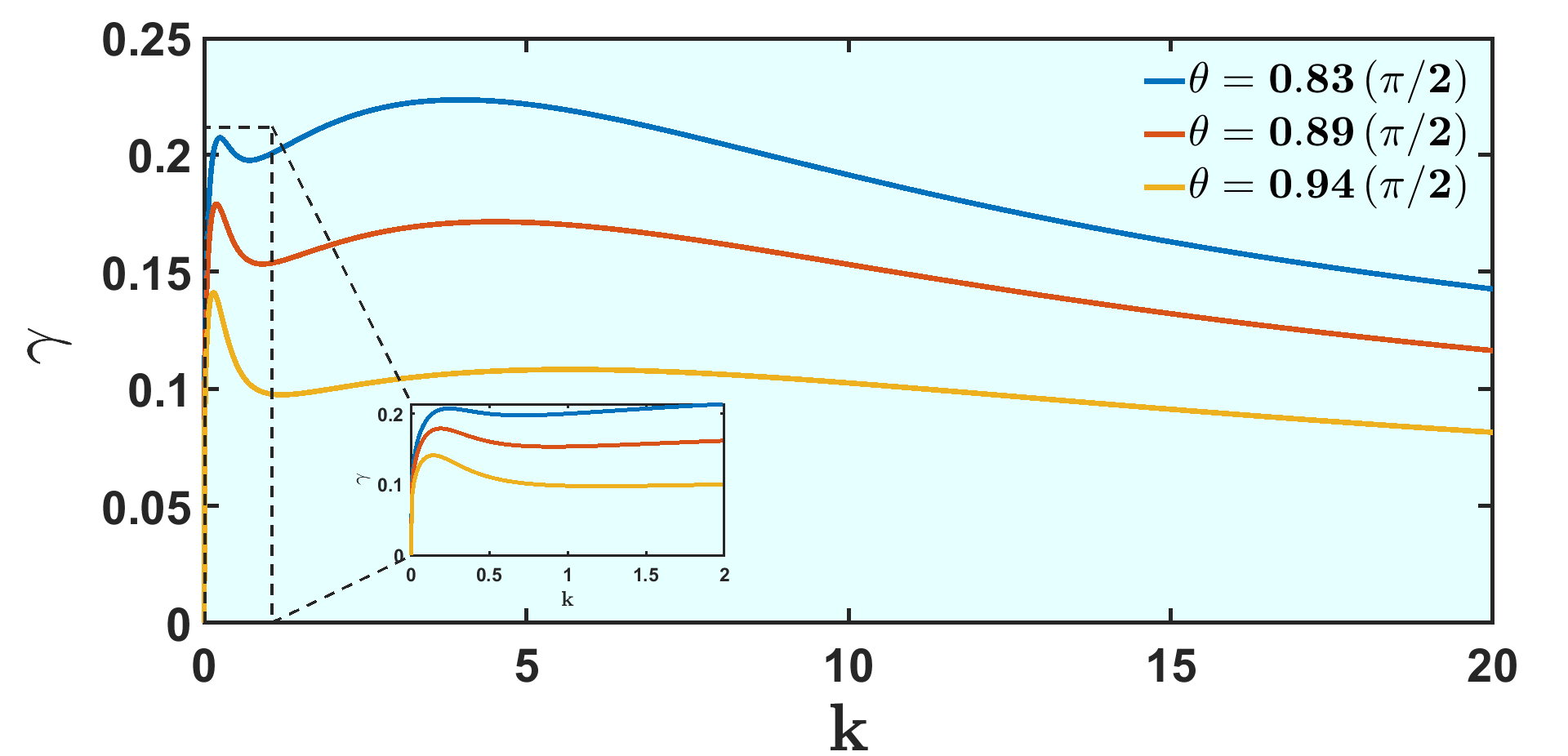}
    \caption{Behaviour of the growth rate for $\omega_{-+}$, for three representative values of $\theta$ ($U = 0.2, \alpha_{YM} = 3.0$). Note that for $k << \kappa_{YM}$, an additional peak emerges.}
    \label{fig:Growth-YM-M1-zoom}
\end{figure}

\begin{figure}
    \centering
    \includegraphics[width=0.80\textwidth]{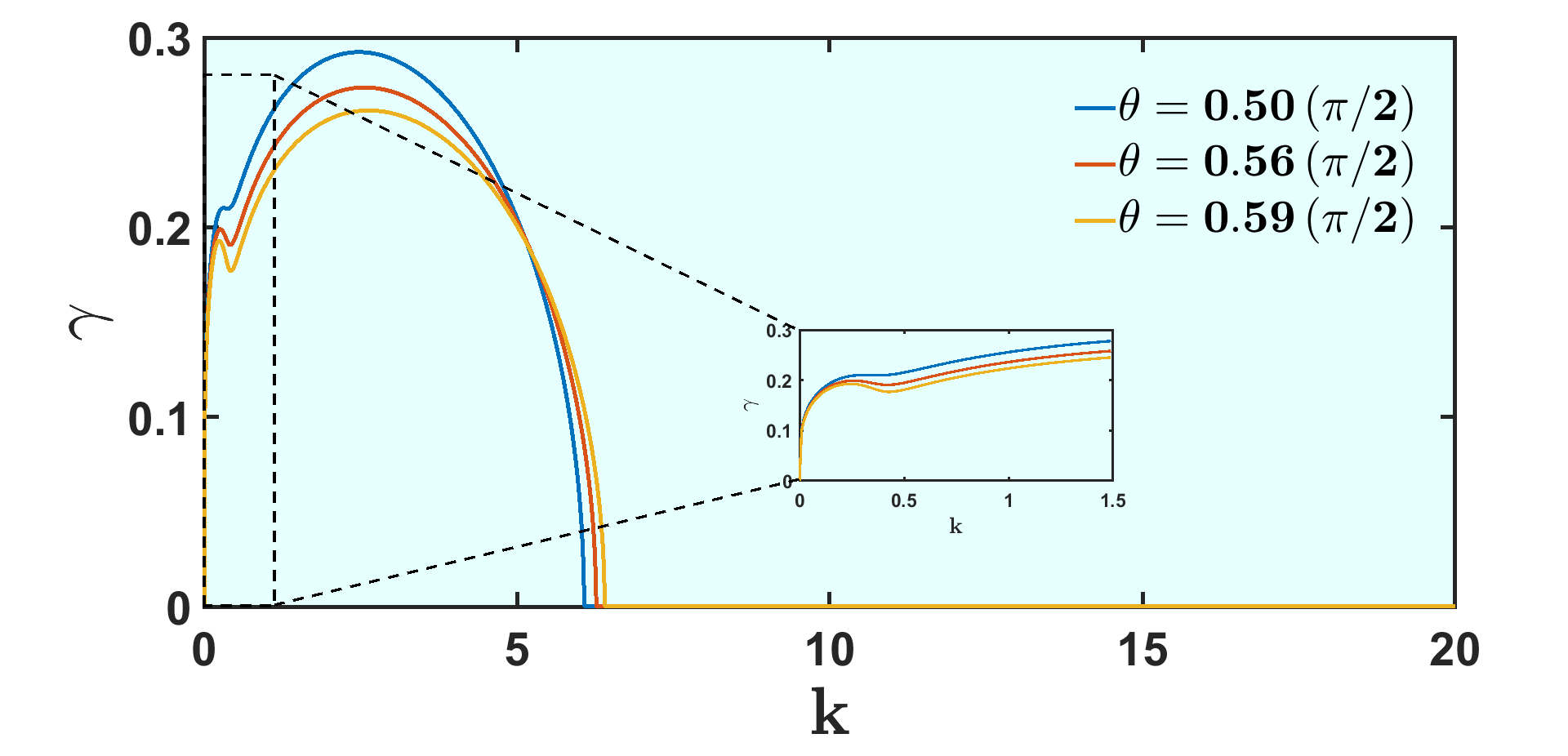}
    \caption{Behaviour of the growth rate for $\omega_{+-}$, for three representative values of $\theta$ ($U = 0.2, \alpha_{YM} = 3.0$). Note that for $k << \kappa_{YM}$, an additional peak emerges. }
    \label{fig:Growth-YM-M2-zoom}
\end{figure}

When the value of $\theta$ is close to $\pi/2$ an interesting feature emerges. This can be observed in Fig.\ref{fig:Growth-YM-M1-zoom} and Fig.\ref{fig:Growth-YM-M2-zoom}. The growth rate acquires an additional peak at lower values of $k$ for both the growing modes. We will discuss the behaviour of this peak and the physics associated with it in the next subsection (\ref{sec:kink-explain}).
\noindent
Table\ref{tab:Property-gen-growth} summarises the findings described above.

\begin{table}[h!]
    \centering
    \begin{tabular}{|m{1.3cm}|m{0.9cm}|m{3.2cm}|m{2.0cm}|m{2.0cm}|m{3.0cm}|}
        \multicolumn{6}{c}{\textbf{Table 3}} \\ [1.0ex]
        \hline
        \textbf{Solution} & \textbf{Mode} & \textbf{Nature of \newline the Mode} & $\mathbf{\gamma_{max}(\theta)}$  & $\mathbf{k_{c}(\theta)}$ & \textbf{Presence of an additional peak} \\ [1.5ex]
        \hline
        $\omega_{++}$ & $\mathcal{E}_R$ & Purely Propagating & Not \newline Applicable & Not \newline Applicable & Not \newline Applicable \\ [1.5ex]
        \hline
        $\omega_{-+}$ &$\mathcal{E}_R$ & Purely Growing & Decreases \newline with $\theta$& No cut-off & Yes \\ [1.5ex]
        \hline
        $\omega_{+-}$ &$\mathcal{E}_L$ & Conditional Growth \newline+ Propagation  & Decreases\newline with $\theta$ & Increases \newline with $\theta$& Yes \\ [1.5ex]
        \hline
        $\omega_{+-}$ &$\mathcal{E}_L$ & Conditional Growth \newline+ Propagation  & Decreases\newline with $\theta$ & Increases \newline with $\theta$ & Yes \\ [1.5ex]
        \hline
    \end{tabular}
    \caption{Brief summary of the defining features of the modes.}
    \label{tab:Property-gen-growth}
\end{table}
There is also an interesting propagating mode in the system which becomes increasingly relevant near $\theta = \pi/2$ and is responsible for the additional peak in the growth rate that is observed in this regime for the two growing modes. We discuss these issues in detail in the next subsection(\ref{sec:kink-explain}). 

\subsubsection{The propagating mode and the additional peak in the growth rate}\label{sec:kink-explain}
We now analyze the dispersion relation of Eq.(\ref{eq:disp-gen-cond}) near $\theta = \pi/2$ for which 
$k_x$ is very small.  Thus, retaining terms only upto linear  order in  $k_x$  (\ref{eq:disp-gen-cond}) we obtain,
\begin{equation}
    \frac{2 \alpha_{YM}U^3 k_x}{\omega^2\left(\omega^2 - k_y^2 \right)} = \pm 1
\end{equation}
This implies that there is a likelihood of a propagating solution if $(\omega \pm k_y) \rightarrow 0$ as $k_x \rightarrow 0$. We seek a  consistency condition for the same. 

\begin{figure}[h!]
    \centering
    \includegraphics[width=0.70\textwidth]{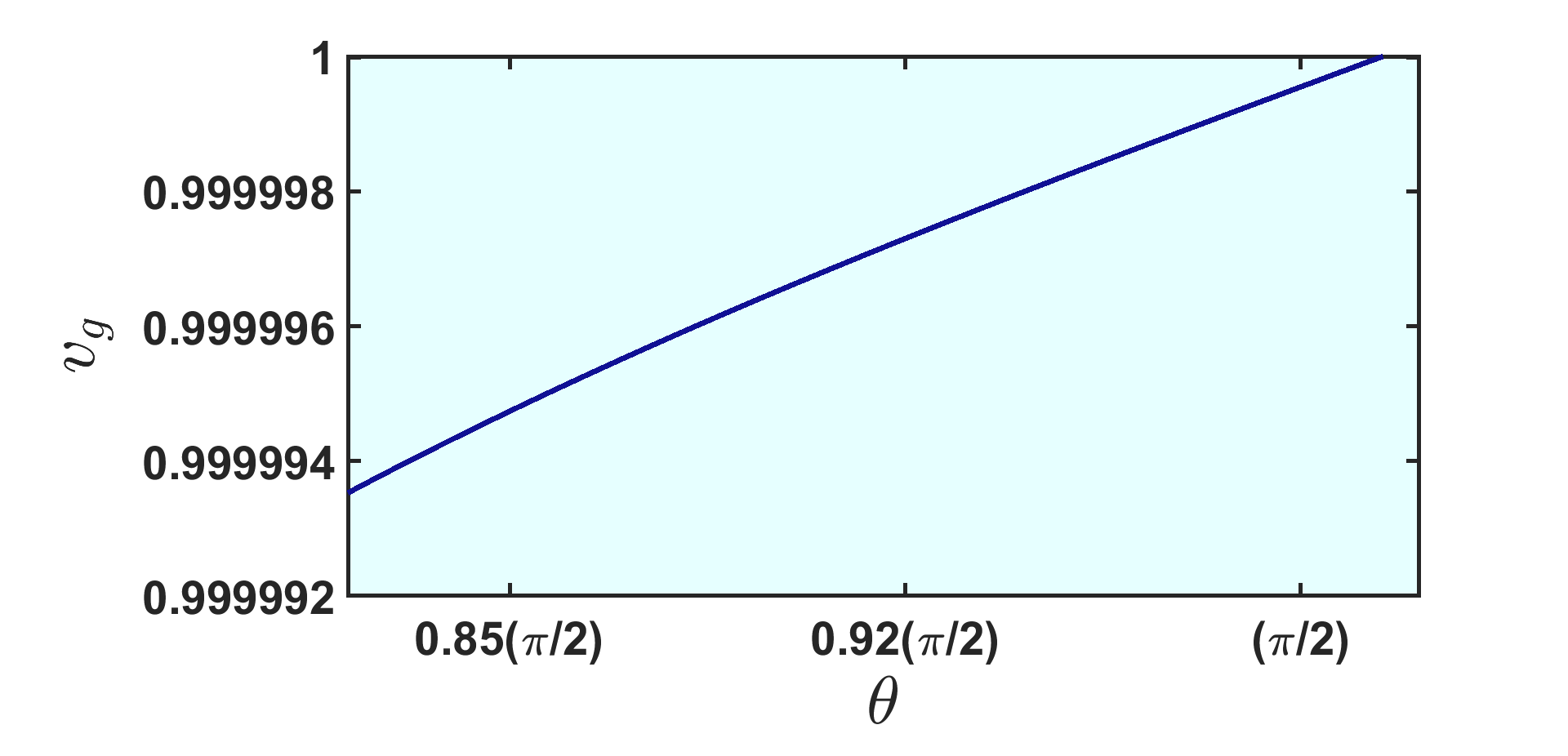}
    \caption{Group velocity for the purely propagating mode described by Eq.(\ref{eq:w-gen}) for $\theta$ close to $\pi/2$, at $k = 15$. Note that $v_g \rightarrow c$ as $\theta \rightarrow \pi/2$.}
    \label{fig:vel}
\end{figure}

Using continuity the consistency condition may be seen to be given by 
\begin{equation}
    k_y^3 = \alpha_{YM} U^3
    \label{eq:mono}
\end{equation}
Thus, near   $\theta = \pi/2$, we get a light-like mode propagating nearly parallel to  $\hat{y}$ with the wavenumber  $k$ given by (\ref{eq:mono}). 
The overall group velocity of this mode approaches $c$ as  $\theta \rightarrow \pi/2$ as shown in Fig.(\ref{fig:vel}). This is a purely  electromagnetic mode and couples to the electric field fluctuations along the 
$\hat{x}$ direction only. 
It is thus important to note that the Y-M plasma permits the vacuum-like propagation of electromagnetic mode at a very specific value of the wavevector with no inherent shielding of plasma. The specific value of the wave vector for which such a vacuum-like propagation happens scales linearly with $U$ and $\alpha_{YM}^{1/3}$. 

\begin{figure}[h!]
    \centering
    \includegraphics[width=0.70\textwidth]{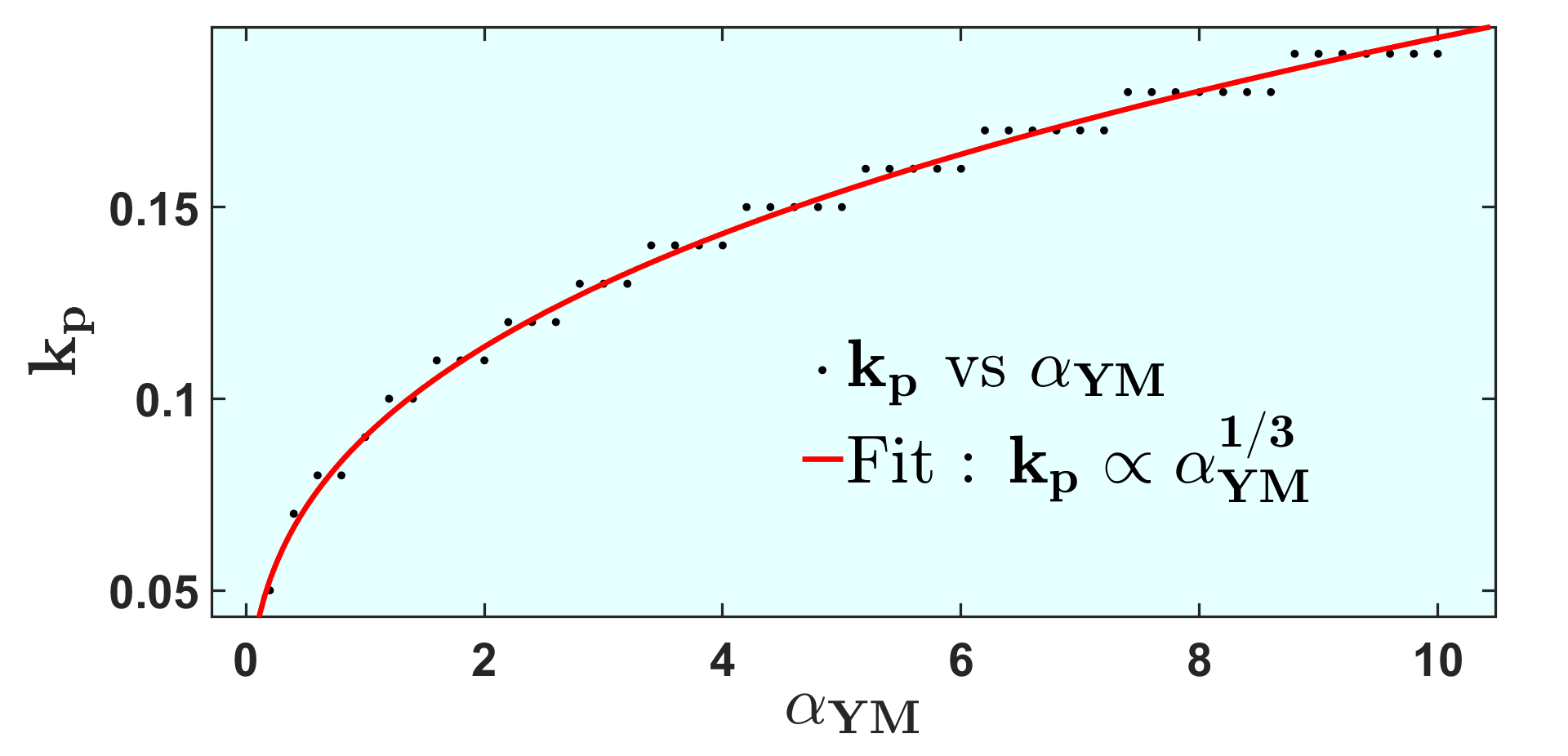}
    \caption{Dependence of $k_p$ on $\alpha_{YM}$ ($\theta = 0.94(\pi/2)$, $U = 0.2$).}
    \label{fig: Kink-explanation1}
\end{figure}

\begin{figure}[h!]
    \centering
    \includegraphics[width=0.70\textwidth]{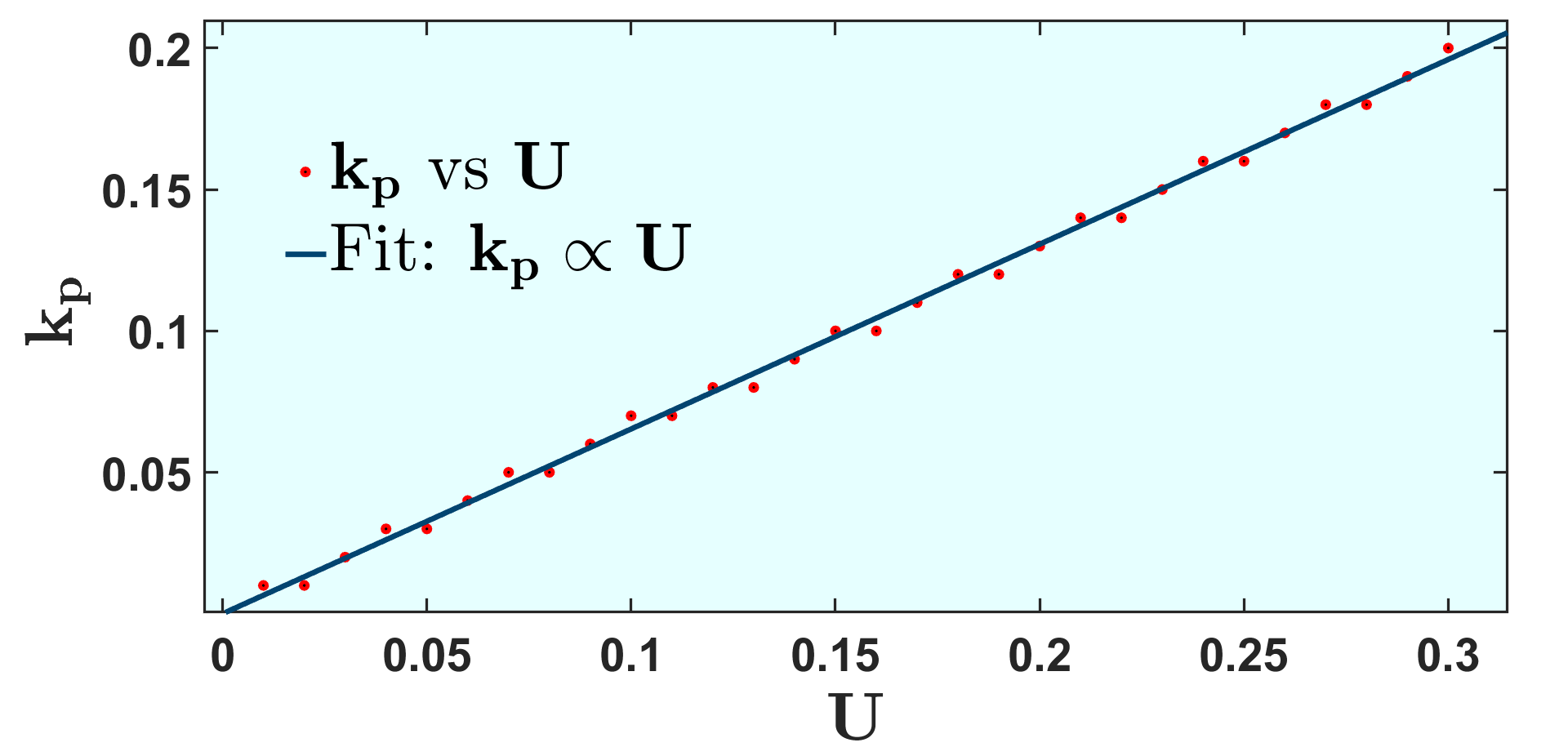}
    \caption{Dependence of $k_p$ on $U$ ($\theta = 0.94(\pi/2)$,$\alpha_{YM} = 3.0$).}
    \label{fig: Kink-explanation2}
\end{figure}

The appearance of the additional peak in the growth rate in Figs. \ref{fig:Growth-YM-M1-zoom},\ref{fig:Growth-YM-M2-zoom} and their inset can now be understood on the basis of this particular radiative mode. A non-monotonicity in any plot signifies the presence of competing physics. Here the free energy available from the flowing plasma also gets coupled with the propagating mode which extracts energy from the system and radiates away. Thereby decreasing the availability of free energy near a certain domain of the spectrum for growth.  
It should be noted that the eigenvectors of the growing modes and the propagating radiative mode are not the usual normal modes. There is thus a coupling of energy between the two kinds of modes even in the linear regime.  The influence of the radiative mode on the growth rate as envisaged above has been conclusively demonstrated in Figs.(\ref{fig: Kink-explanation1}) and (\ref{fig: Kink-explanation2}). The plot of   $k = k_{p}$ (which is a wave vector at which the additional peak in the growth rate appears) as a function of $\alpha_{YM}$ and $U$ in these figures clearly shows that $k_{p}$ scales as  $\alpha_{YM}^{1/3}$ and $U$ which is similar to wave vector Eq.(\ref{eq:mono}) of the radiative mode.


\section{Comparison between the  Abelian and nonabelian growth rates}\label{section:compare}
\begin{figure}[h!]
    \centering
    \includegraphics[width=0.70\textwidth]{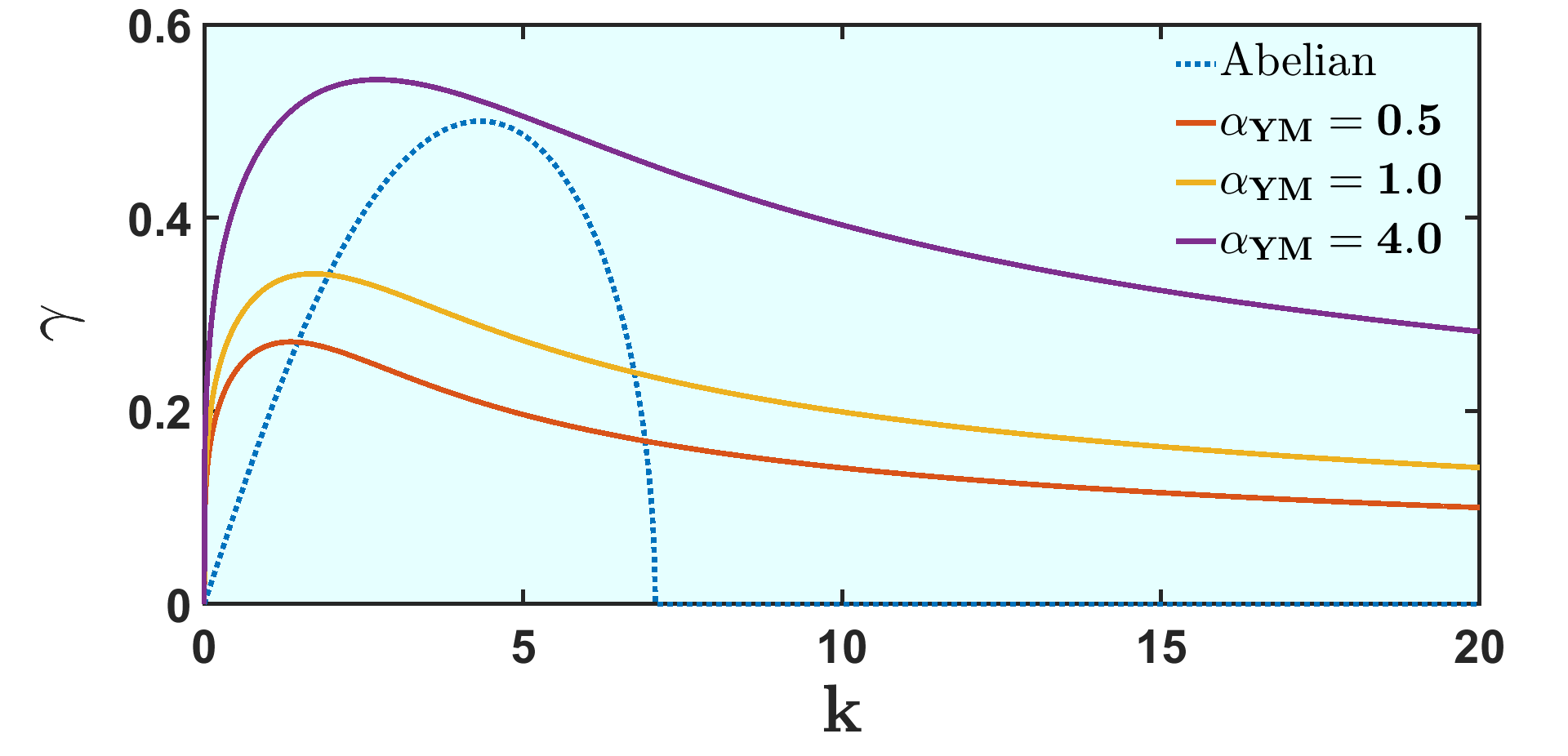}
    \caption{Comparison of the growth rate of the mode $\omega_{-+}$ with the abelian mode (dashed line) described by (Eq.\ref{eq:two-stream}), for three representative values of $\alpha_{YM}$.}
    \label{fig:CompareTheta01}
\end{figure}

\begin{figure}[h!]
    \centering
    \includegraphics[width=0.70\textwidth]{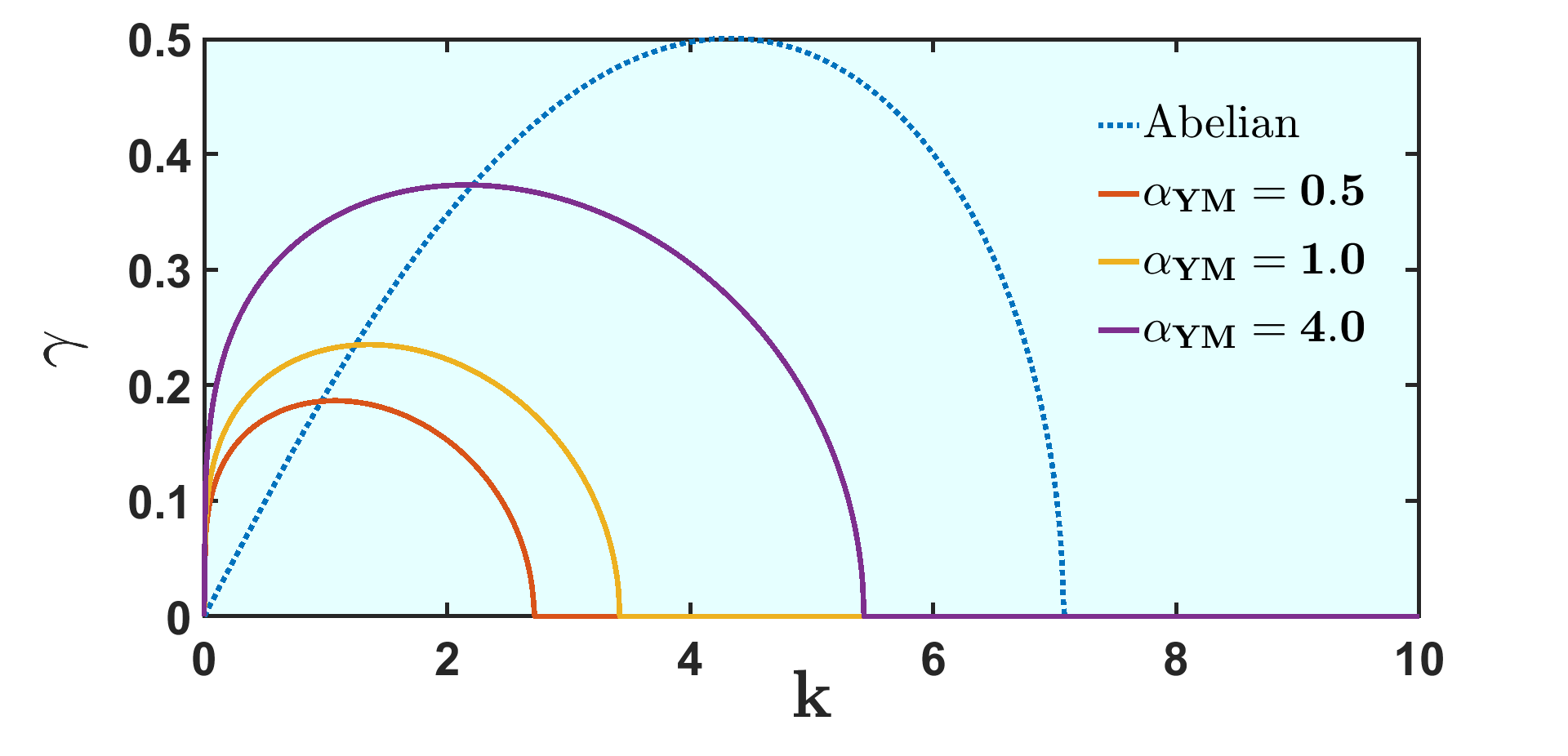}
    \caption{Comparison of the growth rate of the mode $\omega_{+-}$ with the abelian mode (dashed line) described by  (Eq.\ref{eq:two-stream}), for three representative values of $\alpha_{YM}$.}
    \label{fig:CompareTheta02}
\end{figure}

It is important, from an observational perspective, to compare the relative strengths of the growth rates of the nonabelian modes with the conventional abelian ones. One general feature is that at $\theta =0$, the two stream-like nonabelian modes belonging to $\omega_{- +}$ dominate over the abelian counterparts. In fact, by tuning $\alpha_{YM}$ to higher values, it may be made dominant for all $k$ values. More details can be gleaned from Figs. \ref{fig:CompareTheta01} and \ref{fig:CompareTheta02} from which we note that for $k$ small, the growth rates of the nonabelian modes belonging to $\omega_{+ -}$ and $\omega_{- -}$ dominate over that of the abelian mode in the region of the parameter space which we have studied.


\section{Summary and Conclusion}\label{section:Summary and Conclusion}
In this paper, we have studied a Yang-Mills system consisting of two counterflowing fluids of opposite charges and studied the dynamics of instabilities in them with a view to contrast that with the corresponding ED plasma.  
 We find that the nonabelian effects do not present themselves merely as corrections to the familiar ED-like abelian modes,  but also yield an entirely new set of modes which bring out the nonabelian dynamics, plasma dynamics and the effects of flow all at once. We believe that these modes which are exclusive to YM dynamics can have significant implications in understanding the physics of  QGP which are produced in heavy ion collisions. The purely propagating modes may be expected to leave their imprint on the resultant particle jets. Similarly, growths in instability may lead to asymmetries in the angular distributions of the hadrons which are ultimately detected. Any prediction of what they will be would be too premature since we have not accounted for the intrinsic nonlinearity in YM equations.  Furthermore, for any reasonable comparison, it is also necessary to consider the gauge group $SU(3)$ which would introduce its own additional features. These studies will be taken up separately.\\
 
\section*{Credit authorship contribution statement}
\noindent
\textbf{SB}: Numerical work, Analysis, Preparation of Manuscript.
 \textbf{AD}: Conceptualization, Analysis, Preparation of Manuscript. \textbf{VR}: Conceptualization, Analysis, Preparation of Manuscript. \textbf{BP}:  Numerical work, Analysis.

\section*{Declaration of competing interest}
\noindent
The authors do not present any kind of conflict of interest.
\section*{Data availability}
\noindent
No data was used for the research described in the article.
\section*{Acknowledgement}
\noindent
SB wishes to acknowledge the Department of Science and Technology (DST), Govt. of India for providing Inspire Fellowship in support of PhD program. AD acknowledges support from the Science and Engineering Board (SERB)  core grants CRG 2018/000624 and CRG/2022/002782 as well as  J C Bose Fellowship grant JCB/2017/000055, Department of Science and Technology, Government of India.

\bibliographystyle{ieeetr}
\bibliography{reference}


\end{document}